\documentclass[twocolumn,english,aps,pra,longbibliography,superscriptaddress, floatfix]{revtex4-1}
\usepackage{blindtext}
\usepackage[margin=2cm]{geometry}
\usepackage{braket}
\usepackage{comment}
\usepackage{amsmath}
\usepackage{amssymb}
\usepackage{xcolor}
\usepackage{lipsum}
\usepackage{graphicx}
\usepackage[qm]{qcircuit}

\usepackage[%  
    colorlinks=true,
    pdfborder={0 0 0},
    linkcolor=red
]{hyperref}
\usepackage{appendix}
\newcommand{\Hstep}{\ensuremath{H_\text{penal}}}

\newcommand{\x}{\ensuremath{\boldsymbol{x}}}

\begin{document}

\title{Inequality constraints in variational quantum circuits with qudits}
\author{Alberto Bottarelli}
\affiliation{Pitaevskii BEC Center, CNR-INO and Dipartimento di Fisica, Università di Trento, I-38123 Trento, Italy}
\affiliation{INFN-TIFPA, Trento Institute for Fundamental Physics and Applications, Trento, Italy}
\author{Sebastian Schmitt}
\affiliation{Honda Research Institute Europe GmbH, Carl-Legien-Str.\ 30, 63073 Offenbach, Germany}
\author{Philipp Hauke}
\affiliation{Pitaevskii BEC Center, CNR-INO and Dipartimento di Fisica, Università di Trento, I-38123 Trento, Italy}
\affiliation{INFN-TIFPA, Trento Institute for Fundamental Physics and Applications, Trento, Italy}
\date{\today}%July 21, 2022}

\begin{abstract}
Quantum optimization is emerging as a prominent candidate for exploiting the capabilities of near-term quantum devices. Many application-relevant optimization tasks require the inclusion of inequality constraints, usually handled by enlarging the Hilbert space through the addition of slack variables.
This approach, however, requires significant additional resources especially when considering multiple constraints.
Here, we study an alternative direct implementation of these constraints within the QAOA algorithm, achieved using qudit-SUM gates, and compare it to the slack variable method generalized to qudits.
We benchmark these approaches on three paradigmatic optimization problems. We find that the direct implementation of the inequality penalties vastly outperforms the slack variables method, especially when studying real-world inspired problems with many constraints. 
Within the direct penalty implementation, a linear energy penalty for unfeasible states outperforms other investigated functional forms, such as the canonical quadratic penalty. 
The proposed approach may thus be an enabling step for approaching realistic industry-scale and fundamental science problems with large numbers of inequality constraints.
\end{abstract}          

\maketitle
%%%%%%%%%%%%%%%%%%%%%%%%%%%%%%%%%%%%%%%%%%%%%%%%%%%%%%%%%%%%%%%%%%%%
\section{Introduction}

In recent years, quantum optimization \cite{abbas_QuantumOptmization_2023,symons_GuideQuantumOptmization_2023} has emerged as a highly promising field within near term quantum computation, as it may help to solve combinatorial optimization problems prevalent in both physics and industrial applications. Examples include Max-Cut, $k$-sat, various portfolio management, scheduling, and assignment problems. 
Numerous quantum algorithms have been developed to tackle these optimization tasks, leveraging different concepts such as adiabaticity, employed in quantum annealing \cite{hauke2020annealing,yarkoni2021}, or the variational principle, at the basis of the Variational Quantum Eigensolver (VQE) \cite{NISQAlgo,cerezo2021variational,tilly_VQE_review2022,fedorov2022_VQEreview} or Quantum Approximate Optimization Algorithm (QAOA) \cite{farhi2014quantum,blekos2024QAOA_Review}.
These algorithms have in common that they formulate the optimization problem in terms of a cost function $C(\x)$ defined on the boolean cube, where the objective is to find a string $\x\in\{0,1\}^N$ that minimizes or maximizes $C(\x)$.
For realistic applications, a central aspect is given by the presence of (typically  many) constraints. See, for example, Ref.~\cite{limmer_EV_LNS_2023} for a recent example from the electromobility domain. 
These are commonly implemented in quantum optimization routines by adding a penalty function in the cost Hamiltonian, analogous to 
classical optimization approaches.
However, often the constraints are of inequality type, whose realization, typically done in the form of slack variables, requires significant overhead compared to equality constraints.  
The inclusion of inequality constraints thus represents a major bottleneck for realistic quantum-optimization protocols, in particular for their implementations on current noisy intermediate-scale quantum (NISQ)  hardware. 

\begin{figure}
    \centering
    \includegraphics[width = \linewidth]{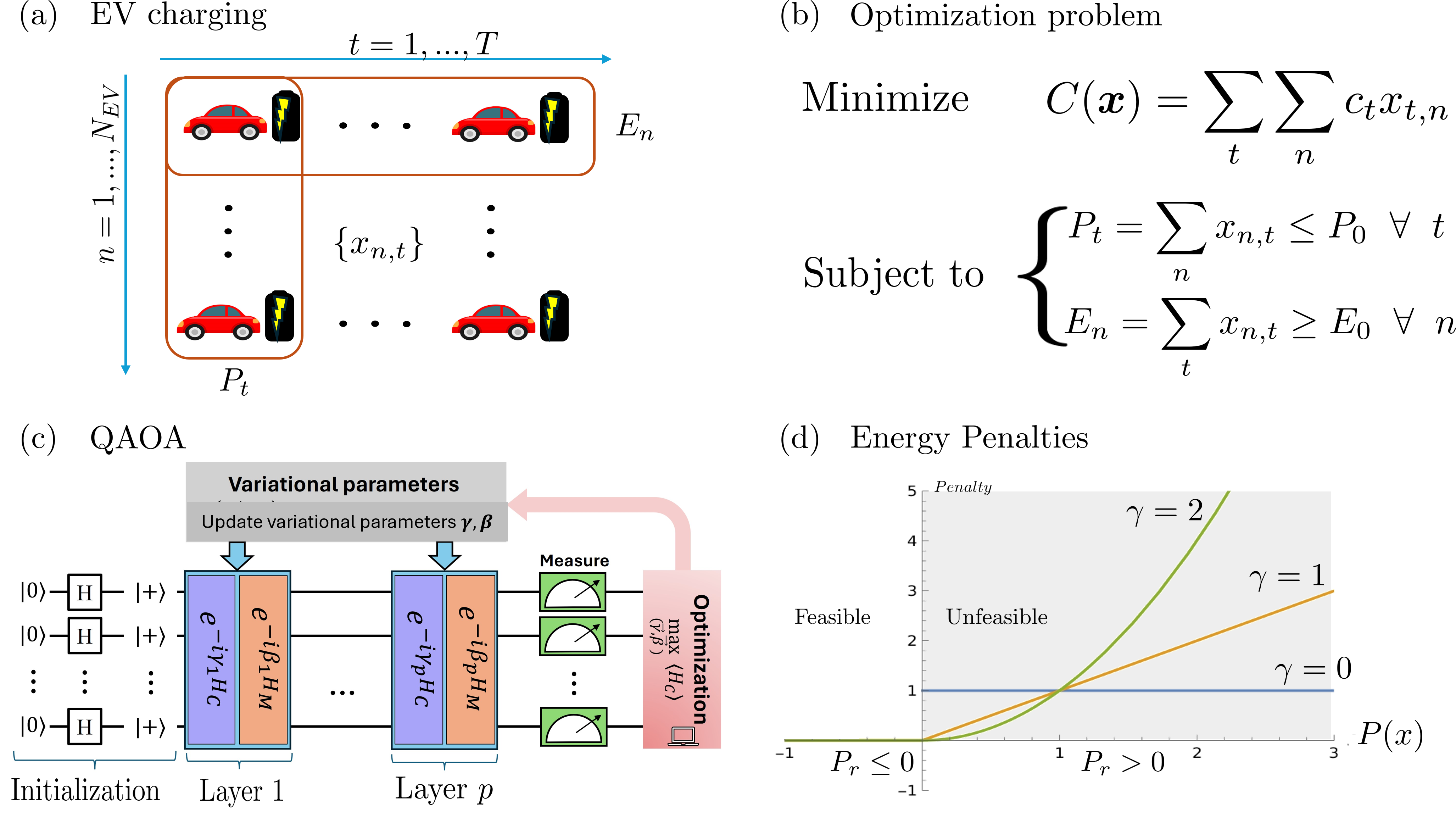}
    \caption{Schematic representation of the aspects of an electric vehicle (EV) charging problem as a typical realistic optimization problem.   (a) Sketch of the problem structure where $N_{EV}$ EVs need to be charged over a period of $T$ time steps. (b) The cost function to minimize and the constraints, which represent maximal charging power at each time step (indicated by the vertical red squares in panel (a))  and minimum charged energy required for each EV (horizontal red squares in (a)).  
    (c) Form of the proposed energy penalty terms that distinguish feasible and unfeasible configurations and thus allow to enforce inequality constraints. %these constraints can be implemented through  (c).
    (d) Schematic representation of the QAOA procedure..
}
    \label{fig: figone}
\end{figure}

Here, we propose an approach based on the direct implementation of the inequality-constraint energy penalty as diagonal unitaries using qudit \texttt{SUM}-gates. 
We discuss the necessary resource scaling in terms of qudit levels and gates and compare it to the standard constraint-handling approach based on slack variables. We show that for a specific class of constraints, we obtain an exponential reduction in the number of gates needed to implement the energy penalty. e illustrate the direct energy penalty on three different problems, a randomly interacting Ising spin model, a constrained state sampling problem, and an electric vehicle (EV) charging problem [see Fig.~\ref{fig: figone}(a,b)], 
and benchmark it against the slack-variable approach using exact numerical simulations of a QAOA protocol [Fig.~\ref{fig: figone}(c)].  
As a further advantage, the functional form of the direct energy penalty can be chosen essentially at will.  
Generically, we find the precise form of the energy penalty [Fig.~\ref{fig: figone}(d)] to have a significant influence on the performance of the quantum optimization protocol. In particular, a linear function outperforms other forms, such as a constant or a quadratic function. 
Furthermore, we find that the proposed direct unitary method performs better than the slack variable state-of-the-art method, especially when multiple constraints have to be considered.  

This enhancement is facilitated by tapping into the rapid recent developments in qudit quantum information processing. Precise and universal control of  quantum systems with $d>2$ levels has been realized on various platforms ranging from trapped ions, over neutral atoms, to superconducting systems \cite{qutrit_experiement2021,ringbauer2022universal,kasper_universal_2022,chi2022programmable,gonzalez2022hardware,ringbauer_entanglement2023,gao2023roleofentanglement,fischer2023universal}. 
By compressing quantum information into higher-dimensional objects, the use of qudits is particularly appealing in  the NISQ-era, where even a constant or polynomial saving of resources can have a significant effect. 
The usefulness of qudits has been discussed for example in the context of combinatorial quantum optimization problems \cite{bravyi2022,deller2023quantum,bottrillQutrit2023,karacsony2023efficient}, where the natural description is in terms of $d$-ary integer variables $ x_i\in \{0,...,d-1\}$ rather than only binary variables. 
Our results illustrate another useful application by using ancilla qudits to implement inequality constraints as a direct energy penalty in the cost function may generate a significant performance advantage for realistic optimization problems. 

The remainder of this paper is organized as follows. In Sec.~\ref{sec: QAOA}, we recap QAOA for qubit and qudit systems. 
In Sec.~\ref{sec: constraints}, we discuss possible ways of imposing constraints on the solutions of the problem. In particular, we compare their resource scaling for the standard approach via slack variables and the method that is the main subject of this study, the inclusion of a nonlinear energy penalty in the cost function.  
Section~\ref{sec: numerics} presents a numerical study of the approaches described in previous sections, showing the beneficial performance of the direct approach to constraints through diagonal unitaries. We present our conclusions in Sec.~\ref{sec: conclusions}. Technical details and explicit circuits used in order to introduce the penalties are reported in the Appendix.
%%%%%%%%%%%%%%%%%%%%%%%%%%%%%%%%%%%%%%%%%%%%%%%%%%%%%%%%%%%%%%%%%%%%
\section{Review of the Quantum Approximate Optimization Algorithm (QAOA)}\label{sec: QAOA}
%%%%%%%%%%%%%%%%%%%%%%%%%%%%%%%%%%%%%%%%%%%%%%%%%%%%%%%%%%%%%%%%%%%%
The quantum approximate optimization algorithm (QAOA)~\cite{blekos2024QAOA_Review,farhi2014quantum, farhi2019quantum} is a hybrid variational quantum algorithm~\cite{NISQAlgo,abbas_QuantumOptmization_2023} designed to solve combinatorial optimization problems. Its schematic is shown in Fig.~\ref{fig: figone} (c). Given a cost function $C(\boldsymbol{x})$ of $N$ classical variables $\boldsymbol{x}=(x_1,...,x_N) $, QAOA aims at finding the solution to 
\begin{equation} \label{eq:minimize}
    \underset{\boldsymbol{x}\in\{0,1\}^N}{\mathrm{min}}C(\boldsymbol{x}),
\end{equation}
yielding the corresponding optimal configuration $\boldsymbol{x}_{\mathrm{min}}$. 
For now, we assume as usual the variables to be binary, $x_i \in \{0,1\}$. 
Extending  this scheme to multi-level qudits is straight-forward and is described below.    
An important class of problems that is usually addressed with QAOA consists of  quadratic unconstrained binary  optimization (QUBO) problems \cite{kochenberger2014unconstrained} whose cost function is expressed as 
\begin{equation}
    C(\boldsymbol{x})= 4\boldsymbol{x}^TJ\boldsymbol{x}=4\sum_{i,j=1}^{N}J_{ij}x_ix_j\,,
\end{equation}
where $J$ is a symmetric and real $N\times N$ cost matrix, and where we included a factor of $4$ for  convenience. 
Such QUBO problems are important for industrial applications, as many paradigmatic problems can be mapped onto such a form~\cite{Lucas2014,yarkoni2021}, as well as for physics questions given their equivalence to Ising Hamiltonians.

In order to solve the problem using quantum computation, one promotes the classical variables to operators acting on a Hilbert space $\mathcal{H}_C=\mathbb{C}^{2N}$. This is usually defined as
\begin{align}
\label{eq:XToqubitMapping}
    x_i\rightarrow \frac{1+\sigma_z^i}{2} \,,
\end{align}
where $\sigma^i_z$ is the $z$-Pauli spin operator for qubit $i$. 
The cost function $C(\boldsymbol{x})$ is promoted to a Hamiltonian $H_C$ by replacing each variable with the corresponding operator, as indicated in Eq.~\eqref{eq:XToqubitMapping}. 
Each state of the computational basis $\{\ket{\boldsymbol{x}}\}$ represents a classical bit string $\boldsymbol{x}$ and $H_C$ is diagonal on this basis with eigenvalues $C(\boldsymbol{x})$:
\begin{equation}
    H_C\ket{\boldsymbol{x}}=C(\boldsymbol{x})\ket{\boldsymbol{x}}.
\end{equation}
The solution of the optimization problem is thus encoded in the ground state of the Ising Hamiltonian 
\begin{equation}
    H_C= \sum_{i,j}J_{ij}\sigma_i^z\sigma_j^z+\sum_{i}h_i\sigma_i^z \,.   
\end{equation}

The QAOA ansatz for finding the ground state of $H_C$ consists in preparing a parametrized trial state 
\begin{align}
\ket{\psi(\boldsymbol{\alpha},\boldsymbol{\beta})}= U(\boldsymbol{\alpha},\boldsymbol{\beta})\ket{\psi_0},\label{eq:qaoa_trialstate}  
\end{align}
generated by alternating parametrized unitaries in the following way:
\begin{align}
   U(\boldsymbol{\alpha},\boldsymbol{\beta})= \prod_{l=1}^{p}e^{i\alpha_l H_C}e^{i\beta_l H_M}\label{eq:qaoa_unitaries}\,.
\end{align}
In the above, $\ket{\psi_0}$ is usually taken to be $\frac{1}{\sqrt{D}}\sum_{\x}\ket{\x}$, with $D=\mathrm{dim}\mathcal{H}_C=2^N$ the Hilbert space dimension and  $\ket{\boldsymbol{x}}$ all the possible states of the computational basis.
A schematic representation of this circuit  is shown in Fig.~\ref{fig: figone}(c).
Apart from a diagonal cost Hamiltonian $H_C$, the state is acted on with a non-diagonal mixing operator $H_M$, which has the role of inducing transitions between the different states. 
It is thus fundamental that the mixing operator and the cost Hamiltonian do not commute, i.e., $[H_C,H_M ] \neq 0 $. Due to the formulation of cost Hamiltonians in terms of Pauli  $\sigma_z$ operators, the most common mixer has the form $H_M = \sum_i \sigma^x_i$, which is also used in this work.
The depth of the ansatz is defined by the integer hyperparameter $p$, also known as the number of layers of the algorithm. 
The expectation value of the cost Hamiltonian is calculated with the trial state of Eq.~\eqref{eq:qaoa_trialstate} as
\begin{equation}
    E(\boldsymbol{\alpha},\boldsymbol{\beta})= \braket{\psi(\boldsymbol{\alpha},\boldsymbol{\beta})|H_C|\psi(\boldsymbol{\alpha},\boldsymbol{\beta})}\,.
\end{equation}
The set of $2p$ variational parameters $(\boldsymbol{\alpha},\boldsymbol{\beta})$ is used to  minimize the expectation value by a classical optimization routine.
Due to the variational principle, this value is ensured to be an 
upper bound to the ground state energy of $H_C$.

It is rather straightforward to generalize QAOA to solve problems formulated with  $d$-ary integer variables $ x_i\in \{0,...,d-1\}$ which are represented by qudits. The local Hilbert space dimension of each operator presenting classical variables becomes $\mathbb{C}^d$ and the Pauli matrices are replaced with angular momentum operators $\{L_x,L_y,L_z\}$ with representation index $\ell=\tfrac{d-1}{2}$. Given a cost function $C(\x)$ defined in terms of such $d-$ary variables, the quantum-mechanical cost Hamiltonian is obtained by replacing each variable as
\begin{align}
    x_i\to L_z^i+\tfrac{d-1}{2}.
\end{align}

 For qudits, the available unitaries are given by the operators of the group $SU(d)$, which has $d^2-1$ generators. To be able to generate all possible unitaries, it is necessary to modify the mixing operator. To the trivial generalization $H_M =\sum_{i=1}^{N}L_x^i$, we add a squeezing operator $\sum_i (L_z^i)^2$.
 It was shown \cite{kasper_universal_2022,giorda_universal_2003} that the set $\{L_z^i,(L_z^i)^2,L_x^i \}$ allows to generate all  possible unitaries of $SU(d)$ by repeated finite rotations of the qudit. 
Therefore,  we use the mixer for qudits in the form 
\begin{align}
    H_M = \sum_i \left(\beta L_x^i + \gamma (L_z^i)^2 \right).
\end{align}
Once the cost Hamiltonian and the mixer are defined, the algorithm works in the same way as the qubit version. 
Below, we will use qudits only for implementing the slack variables, though many application-relevant cost functions can naturally benefit from formulation in terms of qudits \cite{deller2023quantum,bottrillQutrit2023,bravyi2022,wachDRULQudit2022,rocaJeratquditML2023,qudit_KDE_2022,mikelEVqudits2023,vargas_TSP_qudits2021}. 

%%%%%%%%%%%%%%%%%%%%%%%%%%%%%%%%%%%%%%%%%%%%%%%%%%%%%%%%%%%%%%%%%%%%

\section{Constraint handling in QAOA}
\label{sec: constraints}
For many physical and industrial problems, minimizing a cost function as in Eq.~\eqref{eq:minimize} does not necessarily suffice to give the desired solution. Often, the problem is subject to a set of additional constraints that need to be fulfilled. 
Constraints that are expressed as equalities are usually handled in the context of quantum optimization by adding quadratic energy penalties to the problem Hamiltonian, such that unfeasible states are shifted to higher energy~\cite{yarkoni2021,kuroiwa_PenalyVQE_2021}. 
In contrast, inequality constraints are generally harder to include and lead to less efficient encodings. 
These are the subject of this work. 

We focus  on optimization problems of the form 
\begin{align}
\boldsymbol{x}_{\min}&=\underset{\boldsymbol{x}\in\{0,1\}^N}{\mathrm{argmin}} \:C(\boldsymbol{x}) \\
\intertext{ subject to $R$ inequality constraints } 
P_r(\boldsymbol{x})&\leq 0 \:,\quad r=\{1,..., R\}\,,
\label{eq:constraintPr}
\end{align}
where  $r$  labels all the constraints included in the problem and $P_r(\boldsymbol{x})$ is a classical function of the search variables characterizing the constraint.

The current state of the art for incorporating inequality constraints into quantum algorithms consists in converting inequalities into equalities by the use of slack variables and adding corresponding quadratic penalty terms to the Hamiltonian~\cite{yarkoni2021}. 
In contrast to purely qubit based methods, we will employ
qudit slack variables, which avoid the overhead of encoding larger integers into multi binary variables and thus can be seen as a best-case scenario for slack variable approaches (Sec.~\ref{sec:slack}). 
In addition, we propose a novel direct implementation of energy-penalties, which acts only on the unfeasible subspace (Sec.~\ref{sec: penalties}) and allows  to keep the structure of the variational algorithm unchanged (schematics in Fig.~\ref{fig: figone} (c)).
In our numerical benchmarks in the next section, we compare this proposed approach to the qudit-based slack-variable implementation.
For completeness, other possible approaches for inequality-constraint handling that have been discussed in the literature are mentioned in Sec.~\ref{sec: alternatives}.
\subsection{Slack variables as qudit operators}\label{sec:slack}
The standard approach to introduce inequality constraints in QAOA is via the use of slack variables \cite{yarkoni2021}.
Here, we first give a short review of slack variables for classical optimization problems and then discuss how to use qudits in order to encode them in quantum optimization algorithms.
We assume the function $P_r(\x)$ to have $N_r$ possible values, i.e., $P_r(\x)\in \{p_1,...,p_{N_r}\}\,\forall \x$.
Of these values, the first $N^\text{feas}_r$ are negative or zero (i.e., they denote feasible solutions) and $N_r-N^\text{feas}_r$ are greater than zero (i.e., they denote unfeasible solutions). The inequality constraint can be transformed into equality constrains with the introduction of an appropriate slack variable $s_r$ such that 
\begin{align}
    & P_r(\x)\leq 0 \label{eq:in_eq}\\
    \Leftrightarrow\:& P_r(\x) + s_r = 0\,.  \label{eq:in_eq_eq_slack}
\end{align}
The slack variables take non-negative values ($s_r\geq0$) and their range is determined by the values the constraint function attains for feasible configurations such that   only when Eq.~\eqref{eq:in_eq} is satisfied  a value for $s_r$ can be found such that Eq.~\eqref{eq:in_eq_eq_slack}  is satisfied.
Technically, $s_r\in \{-P_r(\x): \forall \text{ feasible } \x\}$. 

The penalized cost function considering all $R$ different constraints can then be expresses as: 
\begin{equation}
    C_\text{slack}(\x)=C(\x)+\sum_{r=1}^{R} \lambda_r (P_r(\x)+s_r)^2\:,
\end{equation}
where one slack variable $s_r$ needs to be introduced for each inequality constraint, and each term gets a penalty factor $\lambda_r>0$.
If the slack variables are encoded into binary degrees of freedom, as is usually done, one requires at least $\sum_r\log_2{N^\text{feas}_r}$ auxiliary binary variables for a given constraint $s_r$ with $N^\text{feas}_r$ allowed values  \cite{nocedal_NumericalOpt_Book2006}. 
Typically, $N_r$ is larger than 2, rendering the associated spatial resource overhead rather costly, especially for current NISQ devices. 

This resource overhead can be reduced by the use of qudits to represent the slack variables. 
If the qudit dimension $d$ matches the number of feasible values $N_r$, the computational basis needs to be extended by one qudit state for each constraint, i.e., slack variable,
\begin{align}
    & \ket{\boldsymbol{x}} 
     \Rightarrow \ket{\boldsymbol{x},\{s_1,\dots, s_{R}\}} \equiv \ket{\boldsymbol{x},\boldsymbol{s} }\,. \label{eq:slack_basis} 
\end{align}

In the quantum formulation, the constraint function as well as the slack variable are represented by operators with the appropriate eigenvalues,
\begin{align}
    (\hat P_r+\hat S_r)\ket{\boldsymbol{x},\boldsymbol{s}} = (P_r(\boldsymbol{x})+s_r) \ket{\boldsymbol{x},\boldsymbol{s}}\,.
\end{align}
With this formulation, feasible configurations of the search variables $\boldsymbol{x}$ can be identified by finding the appropriate basis state where the qudit  slack variable has the correct value to produce a zero eigenvalue in the above equation.

Interestingly, the number of additional dimensions due to the auxiliary qudits (slack variables) is proportional to $N_r^\text{feas}$, the number of  feasible configurations of the corresponding constraint. 
Therefore, the more feasible configurations exist, the larger the dimension of the auxiliary qudit Hilbert space. The dimension of the full Hilbert space including the auxiliary slack qudits  increases  exponentially  with the number of constraints, i.e., $\mathrm{dim}(\mathcal{H}_{\mathrm{slack}}) = 2^N\prod_{r=1}^{R}N^\text{feas}_r$ and 
$\mathcal{H}_\text{slack} = \mathcal{H}_C\otimes  \prod_{r=1}^{R} \mathbb{C}^{N^\text{feas}_r}$.
However, given a feasible configuration $\ket{\x}$, only one state of the extended Hilbert space $\ket{\x,\boldsymbol{s}}$ will represent a feasible total state (the one with eigenvalue $s_r=-P_r(\boldsymbol{x})$), rendering the majority of added quantum states unfeasible. Thus, it can be anticipated that for only lightly constrained problems (that is, with large $N_r^\text{feas}$), the effective optimization problem including the slack variables is more difficult due to the large number of added unfeasible solutions. 

The final form for the constrained Hamiltonian is then
\begin{equation}\label{eq:penalized_ham_slack}
    H_\mathrm{slack} = H_C + \sum_{r=1}^{R}\lambda_r (\hat P_r+ \hat S_r)^2\,.
\end{equation}
The quadratic form of the penalty terms guarantees that it can be made to vanish for all feasible configurations.

Adding qudit slack variables to the system requires modifications to the form of the QAOA-mixer as described at the end of Sec.~\ref{sec: QAOA}. Explictly, we use the following form:
\begin{equation}
    H_M = \beta \bigg(\sum_{i=1}^{N}\sigma_x^i+\sum_{r=1}^{R}L_x^{s_r} \bigg)+ \gamma \sum_{r=1}^{R}(L_z^s)^2 \, ,
\end{equation}  
where both, $\beta$ and $\gamma$ are left as variational parameters.
Apart from this change in the mixing operator, the QAOA is performed as described in Sec.~\ref{sec: QAOA}, just with the cost Hamiltonian given by Eq.~\eqref{eq:penalized_ham_slack}. 

\subsection{ Direct implementation of penalty terms for inequality constraints} \label{sec: penalties}

We propose another possibility for including inequality constraints by using a penalized Hamiltonian of the form
\begin{equation}\label{eq: penalized ham}
    \Hstep = H_C + \sum_{r=1}^{R} \lambda_r \hat G_r\,,
\end{equation}
where $R$ is the total number of constraints and $\lambda_r>0$ are penalty factors for each constraint. 
We introduced the operators $\hat G_r$ whose eigenvalues depend on the corresponding constraint function given in Eq.~\eqref{eq:constraintPr}, namely 

\begin{align}
\hat G_r \ket{\boldsymbol{x}} &=%g(\boldsymbol{x}) \ket{\boldsymbol{x}} =  
g\big(P_r(\boldsymbol{x})\big)\ket{\boldsymbol{x}} \,.
\end{align}

For the function $g(.)$ to penalize only the unfeasible states, it needs to be zero for non-positive arguments. 
To facilitate a gradient towards the feasible subspace, we moreover desire it to be non-decreasing for positive arguments.
These requirements can be achieved by the following choice:
\begin{align}
    g(y) & = y^a \,\Theta(y) \,,
\end{align}
where $\Theta(y)$ is the Heaviside step function and $a\geq0$ is an exponent that can be chosen freely. 
These types of penalty functions are illustrated in Fig.~\ref{fig: figone}(d).
The closest analogy to the slack-variable approach discussed in the previous section is achieved by the quadratic form, $a=2$, but we will see in our numerical analysis in Sec.~\ref{sec: numerics} that $a=1$ produces superior results

Including these penalties, the unitary operator to generate the trial wave function [Eq.~\eqref{eq:qaoa_unitaries}] is updated to 
\begin{align}\label{eq: constrained unitary}
    U(\boldsymbol{\alpha},\boldsymbol{\beta})= \prod_{l=1}^{p}e^{i\alpha_l \Hstep}e^{i\beta_l H_M} 
    \quad \text{with}\\ 
e^{i\alpha_l\Hstep} = e^{i\alpha_l (H_C + \sum_{r=1}^{R} \lambda_r \hat G_r)}\,.\label{eq:phasesGr}
\end{align} 
The rotation generated by the penalizing Hamiltonian puts phases $\sim \alpha_l \lambda_r$ on the unfeasible states, thus permitting the QAOA procedure to select them out.

Since the terms proportional to the constraints are diagonal in the computational basis, it is possible to find  constructions either using Fourier analysis or ancilla registries \cite{welch2014efficient,hadfield2021representation}. 
These methods are defined using only qubit systems and their efficiency is bound by the problem type and instance, and by the properties of the platform chosen for implementation. 

In this work we propose an efficient implementation of diagonal unitaries of the form of Eqs.~\eqref{eq: constrained unitary} and \eqref{eq:phasesGr}. 
In Appendix~\ref{sec: cnots}, we describe the construction which is valid for constraint functions $P(\x)\leq 0$ which only depend on the Hamming weight (i.e., magnetization) $m=\sum_ix_i$ of the state $\ket{\x}$ of $N$ qubits (generalization to summations of $x_i$ and $\overline{x}_i$ is straightforward).
This construction uses one ancilla qudit along with $2(N_r-N^\text{feas}_r)$ qudit \texttt{SUM} gates,
where $N_r-N^\text{feas}_r$  is the number of unfeasible configurations. For the above constraint \eqref{eq:in_eq}, this corresponds to the number of $m$ values with $P(\sum_{i} x_i)>0$, $g(m)>0$.   

In contrast to the slack-variable approach, the ancilla qudit does not enter the cost function, but serves only to imprint appropriate phases according to Eq.~\eqref{eq:phasesGr} onto the trial wavefunction.
The bottleneck of this procedure is the ancilla dimensionality, as it needs to be at least equal to $N+1$. 
However, since the ancilla qudit can be used for many constraints simultaneously, we can expect the proposed method to be advantageous in particular when many constraints are present simultaneously, with each involving only a restricted number of problem  qubits and small to intermediate sets of possible values $P_r$.

\subsection{Alternative approaches for inequality constraints from the literature} \label{sec: alternatives}

For completeness, we also mention other options that have been proposed in the literature. 
For example, one can start from a constrained state and perform the algorithm using only constraint-preserving evolution operators~\cite{fuchs2022constraint,hadfield2019QAOA,baertschiGroverMixer2020}.
The drawback of this method is that usually such constraint-preserving operators are difficult to construct and need to be designed for each specific problem.
Post-selection of feasible solutions~\cite{diezvalle2023multiobj-constraints} can also be used, but depending on the problem this may be very inefficient as finding feasible solutions at all can be difficult, especially for problems where several disconnected domains of feasible solutions exist.   
Inequalities can also be enforced by mid-circuit projection onto the feasible subspace through measurements of the constraint operators~\cite{herman2023constrained}. 
However, these projection operators are problem-specific and can be challenging to implement, while the necessary measurements contribute to both algorithmic and qubit overhead. 

Another recent approach utilizes the augmented Lagrangian formulation of constrained optimization problems~\cite{djidjev_Lagrangian_constrains_2023}, which also introduces energy penalties into the optimization cost function.
However, this approach is only effective for problems where the constraints are active for the optimal solution, i.e., when the optimal solutions lie right at the boundary between feasible and unfeasible solutions such that the constraints become effective equality constraints.
%%%%%%%%%%%%%%%%%%%%%%%%%%%%%%%%%%%%%%%%%%%%%
\section{Numerical benchmarks} \label{sec: numerics}
In this section, we show numerical results for QAOA performed  in different scenarios where inequality constraints play a role. First, we analyze the ability of QAOA to obtain feasible solutions for a generic random spin model~\cite{sauerwein2023engineering}. Then we show how to construct an initial constrained state for warm-starting the QAOA procedure. Finally, we test the performance on the industry-relevant problem of electric vehicle (EV) charging~\cite{sassi2017electric}, which is subject to multiple constraints. For all of these problems, we compare the performance of the constraint-handling methods described in Secs.~\ref{sec:slack} and~\ref{sec: penalties}. 

We simulate the quantum part of the QAOA procedure using exact state vector simulations. 
To update the parameters defining the trial wavefunctions, we use the Powell classical optimizer from the \texttt{scipy} library \cite{virtanen_scipy_2020}. For each problem instance, we perform $N_\mathrm{runs}=50$ separate runs, where for each run the variational parameters are initialized randomly in the interval $[0,2\pi]$. For each finished QAOA run, we extract the classical solutions of the optimization  problem by sampling $N_S=64$ solutions (measurement shots) from the quantum state $\ket{\psi(\alpha^*,\beta^*)}$. 

\subsection{Metrics} \label{sec: metrics}
We use various figures of merit to estimate the performance of the QAOA protocols under different constraint-handling techniques. 
As first metric, we consider the approximation ratio, which is defined as 
\begin{equation}
    R = \underset{\mathrm{samples}\: s}{\mathrm{min}}\frac{E_{s}-E_0}{|E_0|}\,,
\end{equation}
where $E_s$ with $s=1,..,N_S$ is the energy of a state sampled from the final state and $E_0$ is the feasible state with the lowest energy. This metric is relevant for studying the efficiency of QAOA in solving industrial optimization problems, since in such situations the user cares about the best energy of a single configuration that could be achieved, and not the average over the final state.

A second relevant figure of merit is the success rate, defined as 
\begin{equation}
    r = \frac{\sum_{i=1}^{{N}_{\mathrm{runs}}}X_i}{N_{\mathrm{runs}}}\,,
\end{equation}
where $N_{\mathrm{runs}}$ is the total number of runs and  $X_i=1$ if the algorithm sampled at least one of the (possibly degenerate) optimal states in the $N_S$ samples of the $i-$th run, and $X_i=0$ otherwise.  

A third figure of merit is the total weight of feasible solutions present in the final state,
\begin{align}
    & W= \sum_{\boldsymbol{x}\in \mathrm{feasible}}|c_{\x}(\boldsymbol{\alpha^*,\beta^*})|^2\,,
\end{align}
where the $c_{\x}(\boldsymbol{\alpha^*,\beta^*})$ are the amplitudes in the final QAOA state,  
\begin{align}
    &\ket{\psi(\boldsymbol{\alpha^*,\beta^*})} = \sum_{\x}c_{\x}(\boldsymbol{\alpha^*,\beta^*})\ket{\boldsymbol{x}}\,.
\end{align}
This metric serves as an indicator of the efficiency of the algorithm at sampling feasible solutions.
%%%%%%%%%%%%%%%%%%%%%%%%%%%%%%%%%%%%%%%%%%%%%%%%%%%%%%%%%%%%%%%%%%%%
\subsection{Random spin Hamiltonian}
We start by studying the capabilities of QAOA in finding the ground state for a model of randomly interacting spin-$\tfrac{1}{2}$ in a random longitudinal field. We compare the qudit slack variables approach explained in \ref{sec:slack} to penalties discussed in \ref{sec: penalties}. The cost Hamiltonian is
\begin{equation} \label{eq:skrandom}
    H_C=\sum_{i=1}^N h_i \sigma_i^z+\sum_{ij} J_{ij} \sigma_i^z \sigma_j^z,
\end{equation} 
with independent Gaussian-distributed parameters with zero mean and unit variance, i.e., $J_{ij},h_i\sim\mathcal{N}(0,1)$.

For this Hamiltonian, we consider a single constraint affecting all spins of the form
\begin{align}\label{eq: magnetization constraint}
P(\boldsymbol{\sigma}_z)=\sum_{i=1}^N\sigma^z_i-m_0 \equiv S_{\mathrm{tot}}^z-m_0 \leq 0\:,
\end{align}
which filters out all the states with total spin $S_{\mathrm{tot}}^z = \sum_{i=1}^N\sigma_z^i$ less than a target $m_0\in\{-\tfrac N2,\tfrac N2+1,\dots,\tfrac N2\}$.
For the penalties described in Section \ref{sec: penalties}, we choose the exponent $a$ to be equal to $0$, $1$, or $2$, resulting in the penalty Hamiltonians 
\begin{subequations}\label{eq: hamiltonian penalties}
    \begin{equation}
        H_{a=0}  = \lambda\, \Theta(S_{\mathrm{tot}}^z-m_0)\,,\label{eq:flat} 
    \end{equation}
        \begin{equation}
         H_{a=1} = \lambda\, \Theta(S_{\mathrm{tot}}^z-m_0)\,(S_{\mathrm{tot}}^z-m_0)\,,\label{eq:linear}
    \end{equation}
        \begin{equation}
        H_{a=2} = \lambda\, \Theta(S_{\mathrm{tot}}^z-m_0)\,(S_{\mathrm{tot}}^z-m_0)^2\,.\label{eq:quadratic} 
    \end{equation}
\end{subequations}

When using the penalty term for the slack variables as described in Secs.~\ref{sec:slack}, the complete cost Hamiltonian takes the form
\begin{align}
     &H_{\mathrm{C,slack}}  = H_C +
    \lambda\,( S_{\mathrm{tot}}^z-m_0+\hat S)^2\,,\label{eq:slack_penalty} 
\end{align}
where the slack variable operator $\hat S$ acts on a qudit with dimension $d=\tfrac{N}2 +1 +m_0$.
\begin{figure*}
    \centering
    (a)\hspace*{0.5\linewidth}(b)\\
    \includegraphics[width = 0.47\textwidth]{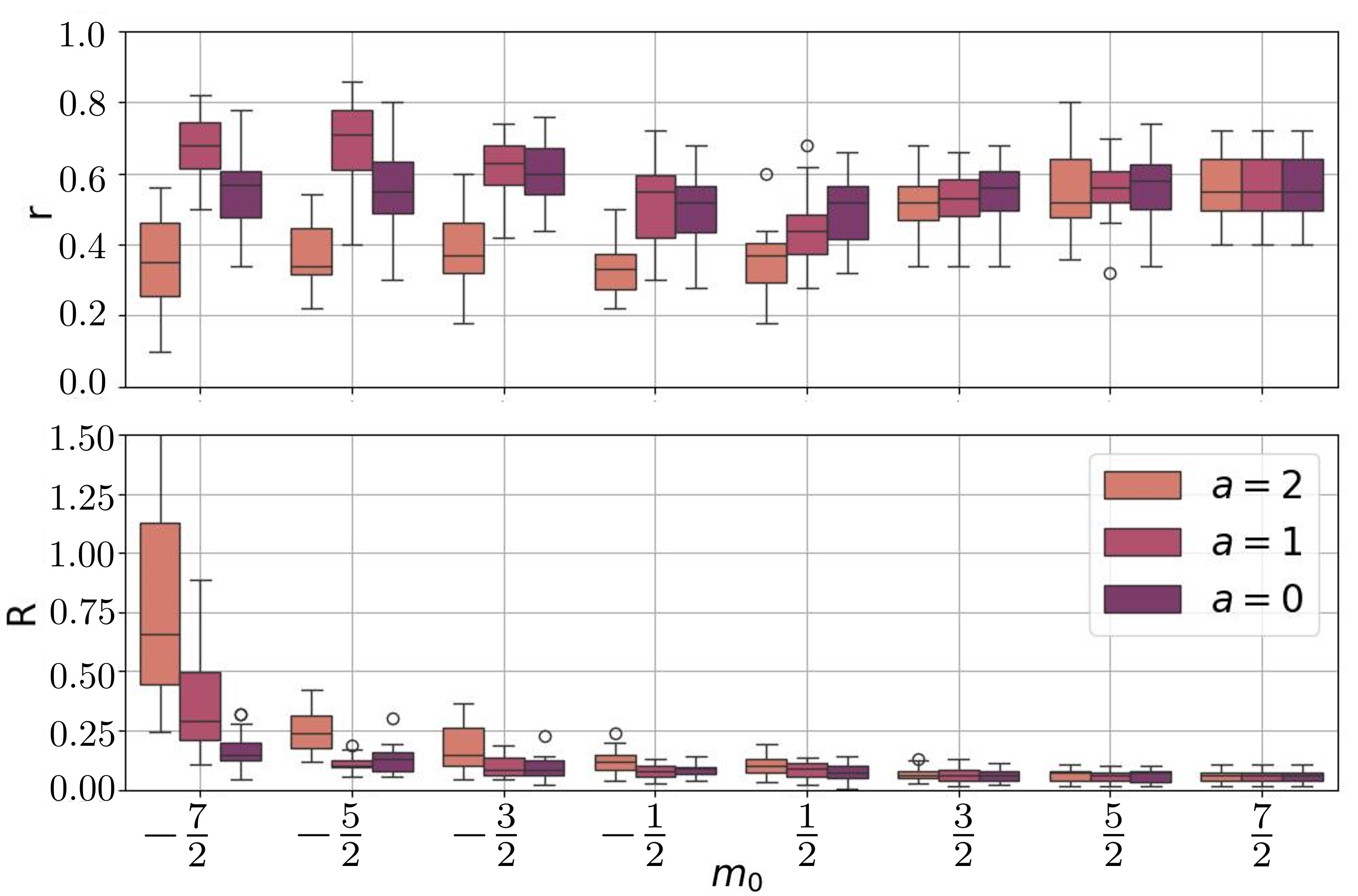}
    \includegraphics[width = 0.465\textwidth]{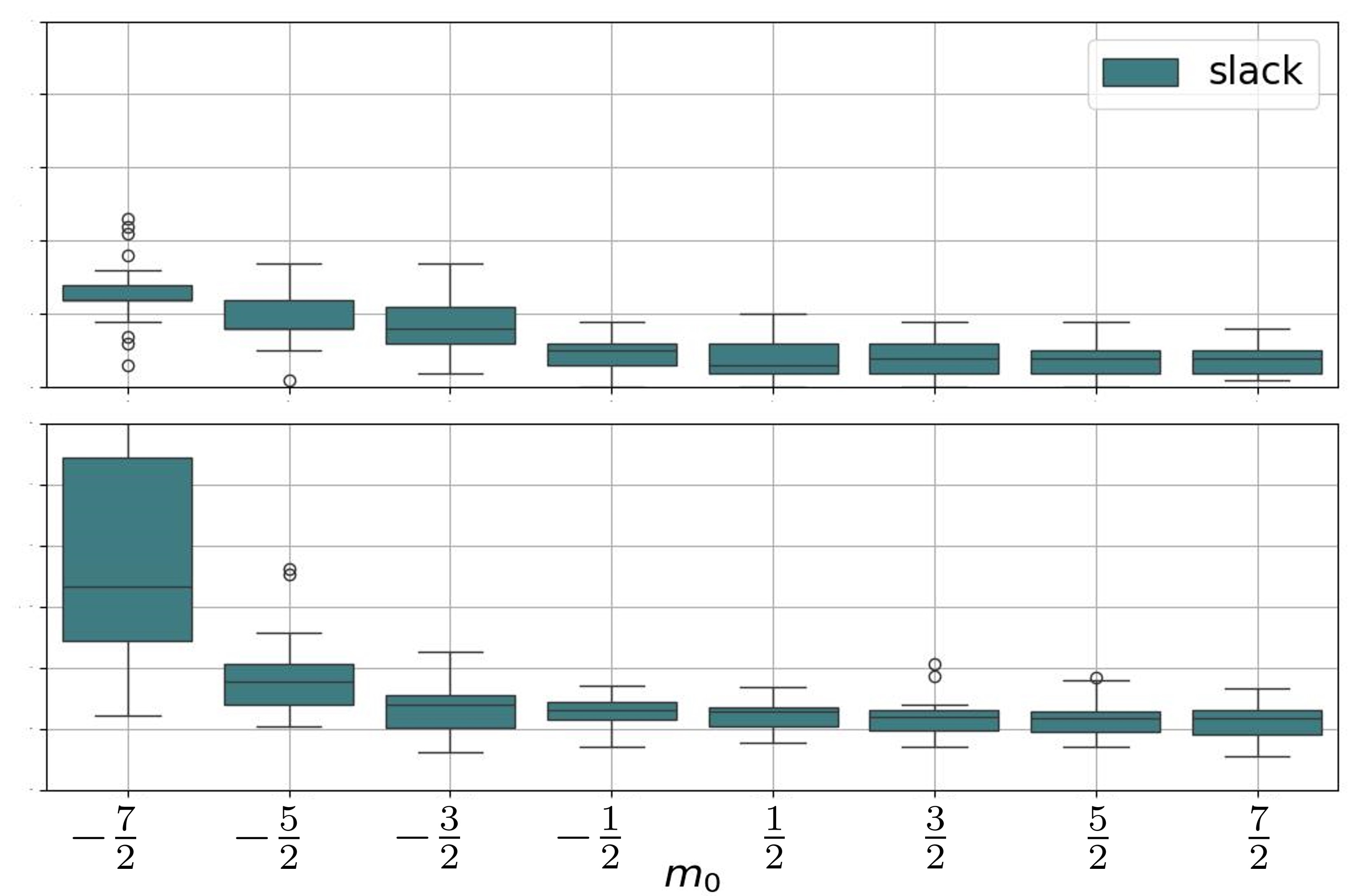}
    \caption{
    Success rate $r$ (top row) and approximation ratio $R$  (bottom row) for constraint QAOA with a single layer ($p=1$) and $N=9$ qubits. The boxplots shows statistics for averages over 20 different instances of the random Hamiltonian of Eq.~\eqref{eq:skrandom}  
    and 50 runs per individual Hamiltonian. The results are shown as a function of the maximally allowed total spin, $m_0$.
    (a) Results from the proposed approach directly implementing the penalty functions of 
    Eqs.~\eqref{eq:flat}$-$\eqref{eq:quadratic}. The linear form ($a=1$) consistently performs best when considering the success rate $r$, while the flat panelty shows best approximation ratio $R$. 
    The results for different $a$ become more similar as the value of $m_0$ increases, since the system becomes increasingly less constrained, reducing the impact of the different forms of penalties. 
    (b) Results from runs using slack variables for handling the inequality constraints. 
    The results clearly show the superiority of the direct penalty approaches of (a) over using slack variables as those achieve much higher success rates and lower approximation ratios. 
    }
    \label{fig: random results}
\end{figure*}

\begin{figure*}
    \centering
    \includegraphics[width = 0.47\textwidth]{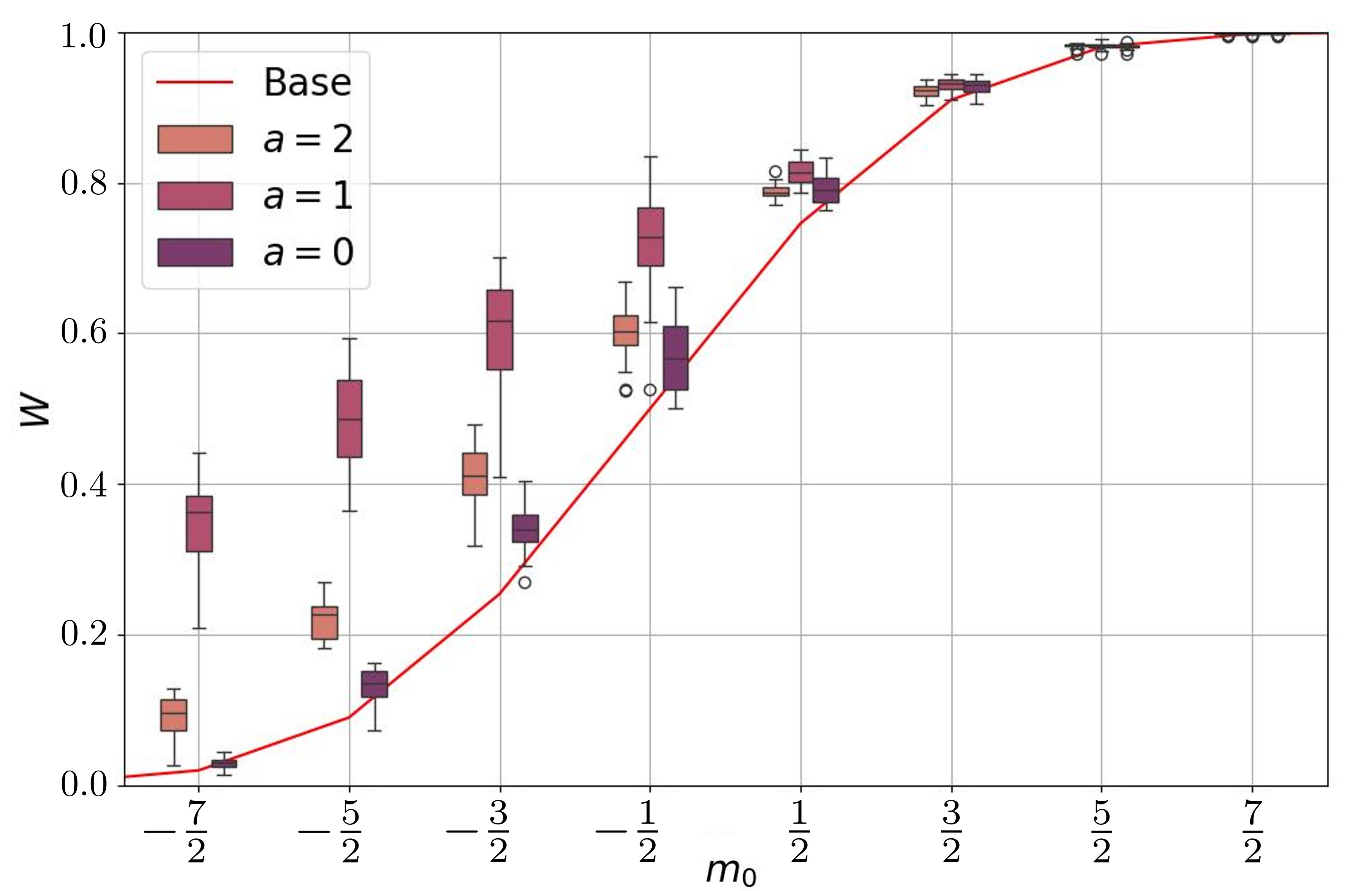}
    \includegraphics[width = 0.457\textwidth]{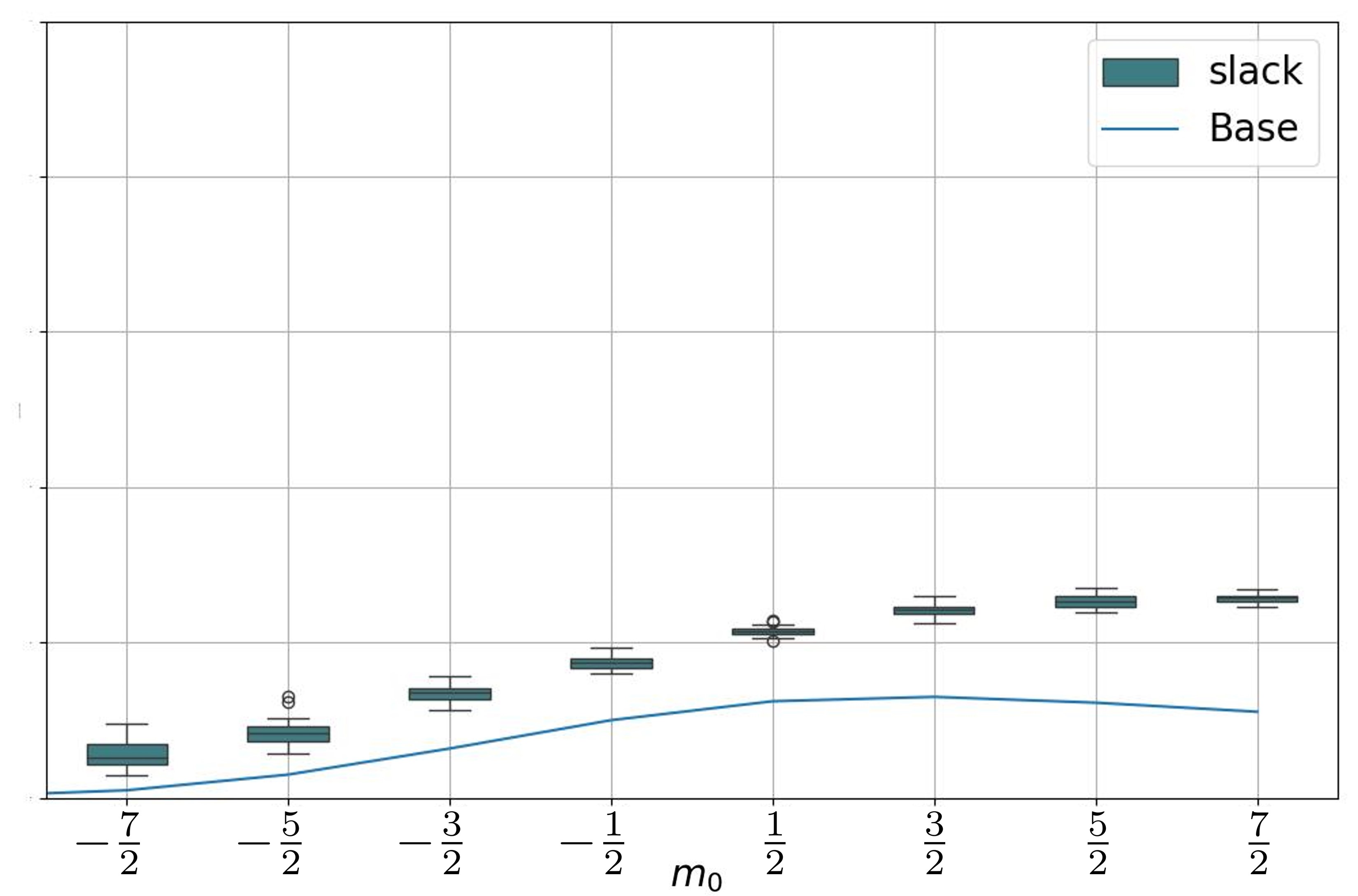}\\
    \caption{Total weight $W$ of all feasible states of  single layer  QAOA with constrained cost function defined in Eqs.~\eqref{eq:flat}$-
    $\eqref{eq:quadratic} (left) and slack variables of Eq.~\eqref{eq:slack_penalty} (right) for the same setup as Fig.~\ref{fig: random results}. The solid lines show the baseline probability of finding a  feasible state from an equal superposition of all basis states. 
    The slack variable approach has consistently lower probability of sampling a feasible configuration from the final state, even if almost all qubit configurations are feasible for larger $m_0$ where it saturates around 25\% while this probability reaches 100\% for the proposed direct penalty method. This trend reflects what we see in Fig.~\ref{fig: random results} and serves as an indicator of the poor performance of slack variables for inequality-constrained optimization.
    }
    \label{fig:feasible subspace weight}
\end{figure*}

We consider a system of $N=9$ qubits, for which we study the metrics defined in Sec.~\ref{sec: metrics} for a $p=1$ layer QAOA as a function of the value of the constrained magnetization $m_0$. 
We generate  $20$ random  realizations of the cost Hamiltonian and run $N_\text{runs}=50 $ simulations for each Hamiltonian. 
All the results shown are obtained for a penalty factor of $\lambda=4$, but higher values for $\lambda$ show similar results.

First, we investigate the efficiency of the different algorithms on the same problems as the dimension of the feasible subspace is increased.
In Figs.~\ref{fig: random results} and \ref{fig:feasible subspace weight}, we compare the results obtained with the proposed method %based on non-linear function penalties  
to the state-of-the-art approach using slack variables.
We see that for all the metrics considered, the performance of the proposed approach is considerably better than that of the implementation with slack variables. 
For example, the success rate $r$ for the penalties of Eqs.~\eqref{eq:flat}$-$\eqref{eq:quadratic} is always significantly higher than for the slack-variable-based approach. Especially when $m_0$ approaches large valuesthe proposed methods converge to $r\gtrsim 0.5$ while for slack variables it converges toward low values around $r\approx 0.1$.
Similarly, the approximation ratio $R$ for slack variables has only small fluctuations around $0.3$ for values of $m_0$ greater than $-2.5$, while for each instance of the approach based on direct penalization it quickly approaches $0$ as $m_0$ increases and the system becomes less constrained. 

Moreover, the performances of the flat, linear, and quadratic penalties can differ considerably.
When considering $r$ as figure of merit, the linear penalty shows the best behavior.
Considering $R$ for highly constrained problems ($m_0\lesssim-1.5$) the best result is obtained with a flat penalty ($a=0$).  
However, this effect is a direct result of the penalty terms in the Hamiltonian. Namely, for these problems the sampled low.energy states will also include unfeasible states, where the actual penalty term contributes to the energy, and consequently linear and quadratic terms will give larger energies in general.

Figure~\ref{fig:feasible subspace weight} shows the total weight of the feasible states as a function of the constraint target value $m_0$, for the three forms of the penalty given in Eqs.~\eqref{eq:flat}$-$\eqref{eq:quadratic} (left panel) as well as for the slack variable approach (right panel).
The continuous lines show the total weight of the feasible states for a uniformly distributed quantum state as a baseline, which is simply given by the fraction of number of feasible states over all states (including qudit-slack variable states in case of slack-variable approaches), i.e., $\text{Base} =\dim(\mathcal{H}_\text{feasible})/ \dim(\mathcal{H}_\mathrm{tot})$.
For the proposed approach, the baseline increases monotonically  from $1/2^N$ for $m_0=-\tfrac N2$ toward $1$ for the unconstrained problem at $m_0=\tfrac N2$.
In contrast, for the slack-variable approach it does not approach unity when increasing  $m_0$; instead, it has a maximum around an intermediate value of $m_0\approx 1.5$ and slightly decreases for larger $m_0$. 
This behavior is due to the changing dimension of the slack variable with $m_0$. On the one hand, the absolute number of feasible states for a given $m_0$ is the same as for the case without slack variables since only one out of all possible values for the slack variable represents a  feasible solutions.
On the other hand, the dimension of the slack variable increases with increasing $m_0$ and thus in case of slack variables the baseline acquires and additional factor $\frac 1d =1 /(m_0+1+\tfrac{N}2)$. 

The results of Fig.~\ref{fig:feasible subspace weight} show that all the tested methods increase the probability of sampling a feasible state over the baseline. However, similarly to the trend of Fig.~\ref{fig: random results}, the proposed direct penalty methods considerably outperform the slack variable approach. 
This effect is most pronounced for the linear penalty function and with increasing $m_0$, i.e., for less constrained problems.
%%%%%%%%%%%%%%%%%%%%%%%%%%%%%%%%%%%%%%%%%%%%%%%%%%%%%%%%%%%%%%%%%%%%
\subsection{Constrained state preparation and sampling of feasible states}

In certain scenarios, the goal of the algorithm may be to sample from a subset of states that fulfil the constraints, without any preference on the states within that subset. Examples include topological models and LGTs ~\cite{kitaev2003fault}, spin ice~\cite{udagawa2021spin}, and the sampling of polymer melts~\cite{micheletti2021polymer,slongo2023quantum}. 
Another scenario is when trying to construct a superposition of many or all feasible states as it is required for the initialization of approaches utilizing constraint preserving mixing operators~\cite{fuchs2022constraint,hadfield2019QAOA,baertschiGroverMixer2020}. 

Motivated by this, we consider a QAOA where the cost function is given only by the penalties encoding the constraint of the form of Eq.~\eqref{eq: magnetization constraint}. 
The QAOA cost Hamiltonian is then of the form of Eqs.~\eqref{eq:flat}$-$\eqref{eq:quadratic} for the proposed approach and of the form of Eq.~\eqref{eq:slack_penalty} for the slack-variable approach.  
We run the QAOA for up to $p=5$ layers, and we do $N_\text{runs}=50$ for each setup (penalty type, number of layers, and constraint value $m_0$).   

\begin{figure}
    \centering
    \includegraphics[width = \linewidth]{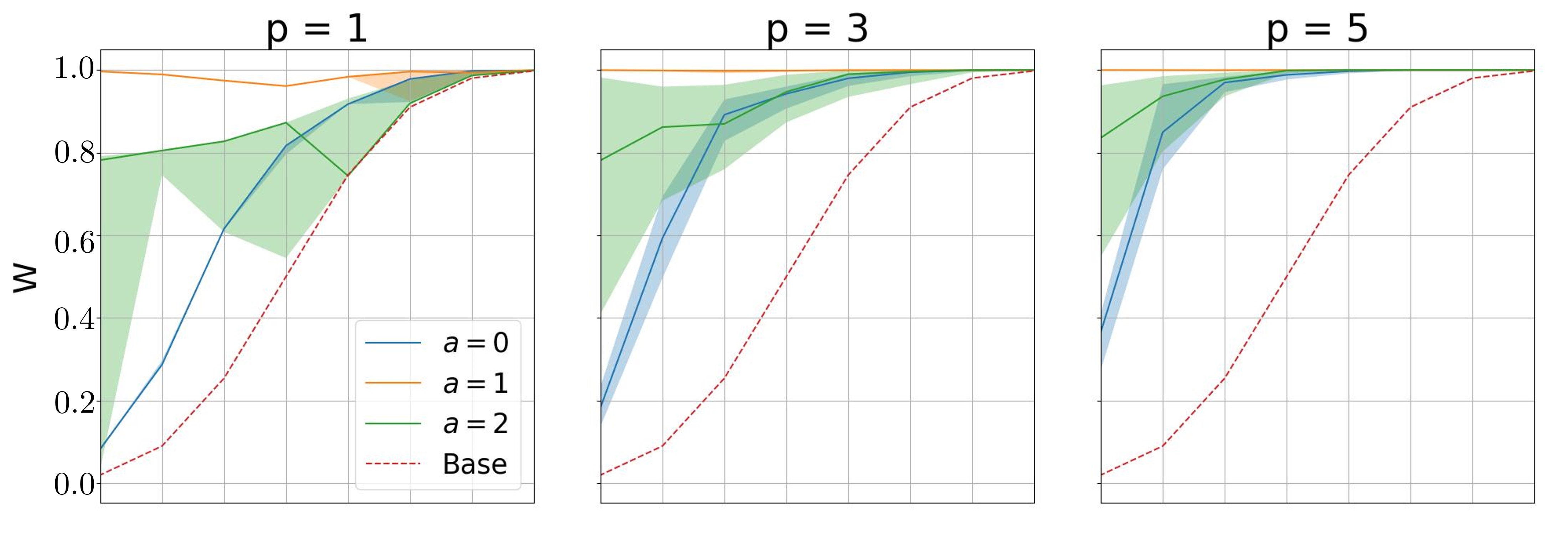}\\
    \includegraphics[width = \linewidth]{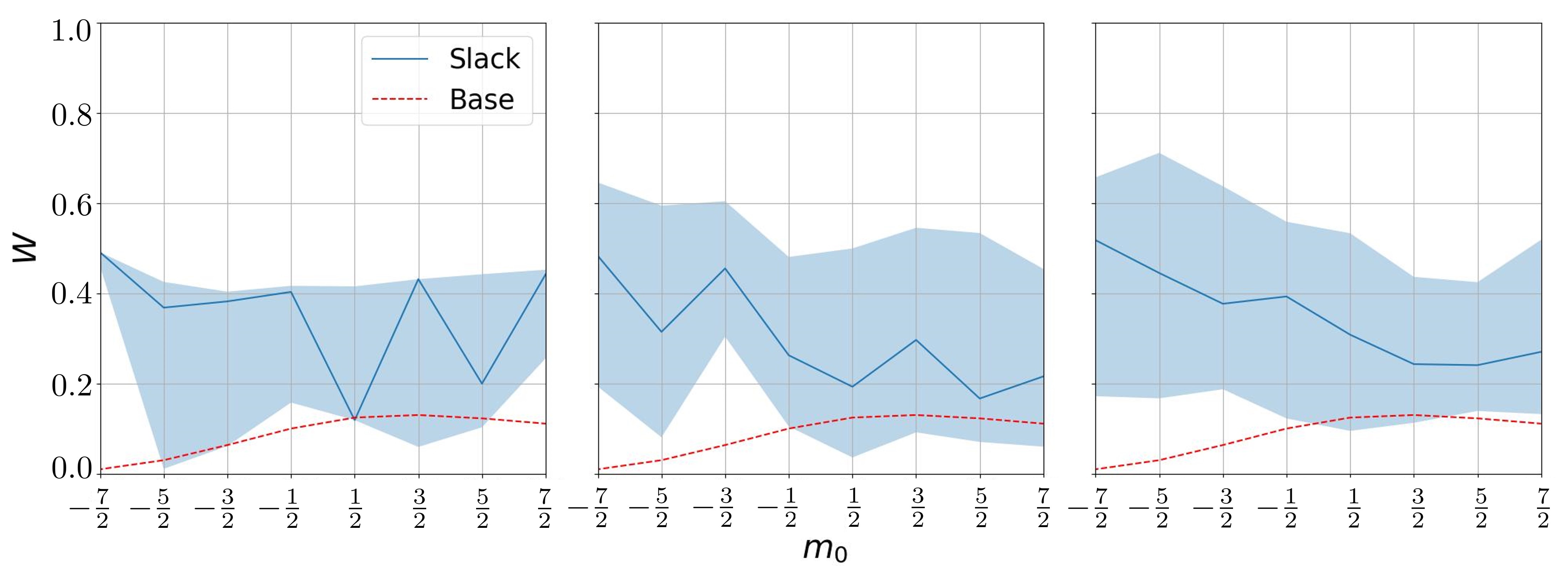}   
    \caption{
    Probability $W$ of sampling a feasible state from the final state of a QAOA with constraint-only cost function, for the direct penalty approach (top row) and slack-variable based approach (bottom), for $p=1$, $3$, and $5$ QAOA layers (left to right panels). 
    The solid lines show median values for the different penalty types and the shaded areas indicate the 20\%-80\% quantiles of the 50 runs performed. 
    The dashed red lines marked as `Base' indicate the baseline probabilities from the equal superposition state of all solutions. 
    The direct penalty functions outperform the slack-variable approach significantly. In particular, the penalty function with $a=1$ gives close to 100\% probability for all constraint values already for only $p=1$ layer.}
    \label{fig:only contraints}
\end{figure}

Figure~\ref{fig:only contraints} shows the total weight of feasible configurations in the final quantum state for numbers of layers $p=1$, $3$, and $5$. 
The linear penalty term considerably outperforms the slack-variable approach as well as the other penalty forms considered, reaching close to the maximum of $W=1$ already after three layers for almost all values of the minimum magnetization $m_0$. 
As in the previous example, all approaches improve as expected over the baseline, given by the weight of the feasible states on the initial state (dashed line). 
Again, for all chosen forms of the penalty function, the proposed direct penalty approach improves significantly more over the baseline than the slack-variable-based approach. 
As it would be expected, the direct approach also achieves overall better performances with increasing number of layers $p$. 
Interestingly such a trend is not observed for the slack-variable based approaches. 

\subsection{EV charging problem}
To make contact with QAOA for solving combinatorics problems of industrial relevance and test the different approaches with multiple constraints, we  evaluate the performance of the proposed approach for an electric vehicle (EV) charging problem. It is schematically represented in Fig.~\ref{fig: figone}(a,b).
The target of this problem is to find the charging schedule for a fleet of EVs, i.e., the charging power for each EV at each time step.     
This problem is naturally formulated in terms of qudit variables~\cite{deller2023quantum} as soon as more than two charging levels (charging or not charging) are considered. 
Importantly in our context, this problems naturally has many inequality constraints since each EV has minimum requirements for the total energy delivered to it while  at each time step the total charging power must not exceed the fuse limits.
It is thus a useful benchmark problem to study how the different approaches perform with the inclusion of multiple constraints.

The cost function of the considered EV charging problem is defined as 
\begin{align} \label{eq: Ev ham}
    C(\boldsymbol{x})&=\sum_{t=1}^{T} c_t \sum_{n=1}^{N_\mathrm{EV}}  \,x_{n,t}\,,
\end{align}
where the variables $x_{n,t}\in\{x_\text{min},\dots,x_\text{max}\}$ characterize the amount of energy charged (or discharged if $x_\text{min}<0$) to EV $n\in\{1,\dots,N_{\mathrm{EV}}\}$ at time step $t\in\{1,\dots,T\}$. 
The (time-dependent) cost for a unit of electric energy is given by the coefficients $c_t$. 
This problem is naturally subject to many constraints.
In particular, we consider the following two types: 
\begin{align}
E_{n}^\mathrm{required}-\sum_{t=1}^T x_{n,t}&  \leq 0 \quad\forall n \,, \label{eq:EV_soc_constr}  \\
\sum_{n=1}^{N_{EV}} x_{n,t} - E^{\mathrm{max}} &\leq 0\quad\forall t \,. \label{eq:EV_pamx_constr} 
\end{align}
The first one, Eq.~\eqref{eq:EV_soc_constr}, reflects the requirement of each EV to obtain a minimal amount of electricity $E_n^\text{required}$ after the charging is finished. 
The second constraint, Eq.~\eqref{eq:EV_pamx_constr}, ensures that the maximal charging energy never exceeds the fuse limits.
These are $N_{EV}+T$ linear constraints in total, where each couples only a specific subset o variables, but taken together they couple all variables with each other. 
This problem is illustrated in Fig.~\ref{fig: figone}.

For demonstration purposes, we consider a rather simple problem instance where we only include two charging power levels, $x_{n,t}\in\{0,1\}$ 
and two vehicles ($N_{EV}=2$) for four time steps ($T=4$). 
The constraints are specified by $E^{\mathrm{max}}=1$, and $E^\text{required}_{0}=E^\text{required}_{1}=2$.
With this setup, the classical combinations that satisfy the constraint are easy to define: each vehicle must be charged at least in two time steps, but the vehicles cannot be charged at the same time. 

The quantum formulation is obtained by the usual replacement of the search variables with spin operators as shown in Eq.~\eqref{eq:XToqubitMapping}. The Hilbert space dimension for this simple example is  dim$(\mathcal{H})=2^{T\cdot N_{EV}}=2^8=256$, out of which only $12$ basis states are feasible. Due to the symmetry with respect to the EVs, the solutions are always doubly degenerate.  
To incorporate the constraints of Eqs.~\eqref{eq:EV_soc_constr} and~\eqref{eq:EV_pamx_constr} using slack variables, we need a total of six auxiliary variables: two qudits for the constraints of Eq.~\eqref{eq:EV_soc_constr} with dimension $d=3$ and four qubits ($d=2$) for those of Eq.~\eqref{eq:EV_pamx_constr}. 
The Hilbert space dimension for the constrained problem is then dim$\mathcal{H}_\mathrm{slack}=2^8\cdot3^2\cdot2^4=9216$, which is two orders of magnitude larger than the Hilbert space of the setup without slack variables. 

We study the performance of QAOA as a function of the number of layers and for different random realizations of the Hamiltonian of Eq.~\eqref{eq: Ev ham}, where the prices are chosen in each instance from a uniform distribution $c_t\sim \mathcal{U}(0,1)$. We run $N_\text{runs}=50$ runs for each problem instance. 
For the approach using direct penalties, we consider $p=1,\dots,5$ layers, while we did only run simulations for up to $p=3$ layers for the slack-variable approach due to its vastly   
larger Hilbert-space dimension and the consequently significantly longer runtimes.

\begin{figure*}
    \centering
    \includegraphics[width = 0.47\textwidth]{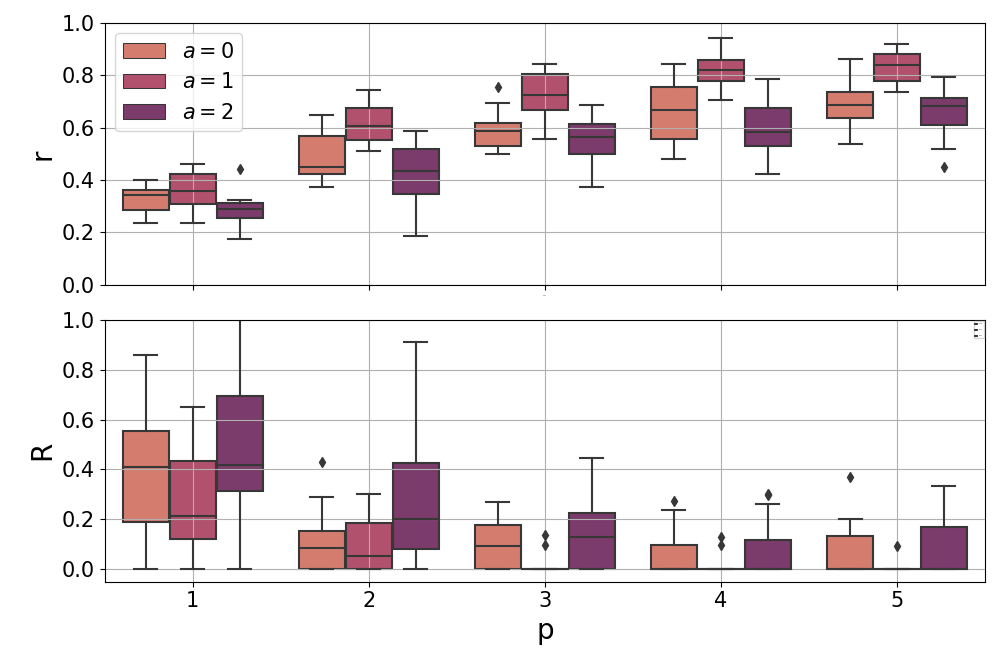}
    \includegraphics[width = 0.47\textwidth]{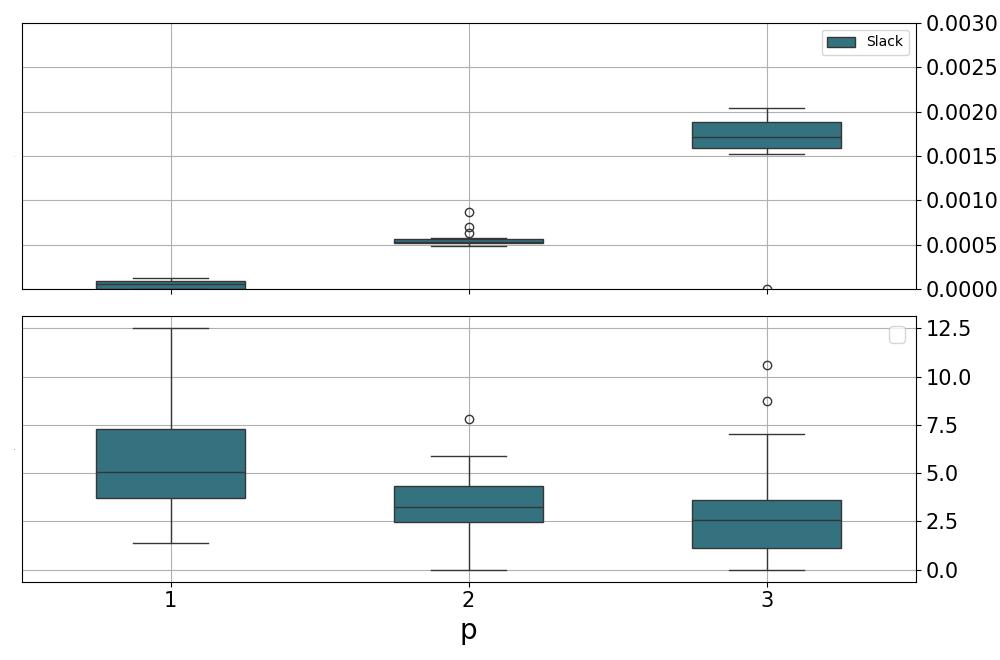}
    \caption{
    Results of the EV charging problem of Eqs.~\eqref{eq: Ev ham} -- \eqref{eq:EV_pamx_constr} for $N_\mathrm{EV}=2$, $T=4$, $E^{\mathrm{max}}=1$, and $E^\text{required}_{n}=2$,  for penalties encoded directly (left) and with slack variables (right) as function of the number of QAOA layers $p$. 
    The approach based on direct penalty terms vastly outperforms the one based on slack-variables (note the different scale of the $y$ axes). 
    In line with the previous findings, the case with linear penalty $a=1$ performs best among the direct penalty approaches. In particular, 
    for $p\geq 3$ the median as well as the $20$-th and the $80$-th quantiles  of the distribution of the approximation ration $R$ all vanish, indicating that an optimal solution is always found for $a=1$.
    }
    \label{fig: results EV}
\end{figure*}
The results, summarized in Fig.~\ref{fig: results EV}, enhances the understanding of the previous benchmarks. 
The approach based on direct penalty Hamiltonians achieves success rates on the order of 30\% already for $p=1$, a value that increases up to 60\%-90\% for $p=5$ layers. 
The approximation ratio also rapidly approaches zero with increasing $p$, especially for the best-performing case of $a=1$.

In stark contrast, the results of the slack variable approach essentially indicate a failure of the method. Although the success rate does increase slightly for a larger number of layers, the overall scale is only around $10^{-3}$. 
Similarly, the approximation ratio is an order of magnitude worse than for the direct penalty approach and essentially indicates that the algorithms never come close to finding the true ground-state energy region.

The reason for this failure is found in the vast increase in Hilbert-space dimension needed to encode all the different constraints. 
For each slack variable, only one of all possible configurations represents a feasible state, which leads to the majority of added states being unfeasible. 
In the above example where there are typically only two feasible configurations for the charging variables $x_{n,t}$, this leads to a reduction of the fraction of feasible solutions from $2/2^8\approx 4\cdot10^{-3}$ without slack variables to $2/(2^8\cdot3^2\cdot2^4)\approx 5\cdot10^{-5}$ with slack variables.
This large increase in search space due to the slack variables makes the corresponding search problem for the QAOA much more difficult.
%%%%%%%%%%%%%%%%%%%%%%%%%%%%%%%%%%%%%%%%%%%%%%%%%%%%%%%%%%%%%%%%%%%%

%%%%%%%%%%%%%%%%%%%%%%%%%%%%%%%%%%%%%%%%%%%%%%%%%%%%%%%%%%%%%%%%%%%%
\section{Conclusion and Outlook}\label{sec: conclusions}
We have presented two approaches of using qudits to include energy penalties for an inequality-constrained QAOA routine: one by directly implementing energy penalties in a diagonal way and another by enlarging the system with ancilla qudits and introducing quadratic slack-variable-based penalty terms. 
We have benchmarked both approaches on three different problems, finding a significant difference in their performance.
Specifically, including constraints directly without relying on additional qudits in the energy function vastly outperforms the slack-variable-based methods. 
Moreover, we find that a linear energy penalty outperforms the constant and quadratic penalty term.

Even if including the constraints through slack variables can lead to some decent results for a single constraint, the performance of this approach drops drastically when including multiple constraints. 
This is the typical case for a large class of combinatorial problems suited for QAOA (e.g., resource allocation problems~\cite{limmer_EV_LNS_2023}). 
Including many slack variables in the problem would be unfeasible for NISQ architectures. 
There, it leads to a drastic increase in required quantum resources that is hard to meet and in turn amplifies the noise in the circuit execution. 
In a fault-tolerant scheme where pure resources and noise are expected to not be the dominating limiting factors,  it still considerably increases the complexity of the circuit, which is known  to be a possible cause for the barren plateaus phenomenon~\cite{larocca2024review}.
Additionally, as shown in this work, the problem of finding low-energy states, i.e., the optimization problem to determine the parameters of the parameterized quantum circuit, is much more difficult to solve due to the vastly increased search space. 

In the future, an interesting direction will be to combine the use of qudits for constraint handling and representation of the cost-function register, in order to maximize the use of available qudit levels in a given machine. Moreover, while qudit encodings of cost functions have been proposed, the potential performance advantages in variational quantum algorithms with respect to the usual qubit formulations need further thorough analysis, in particular also in view of the additional engineering overhead required. 
Our results may also stimulate cross-fertilization with other fields of quantum technologies. 
For example, including constraints in quantum algorithm is of fundamental relevance also for the task of quantum simulation of lattice gauge theories (LGTs)~\cite{banuls2020simulating,halimeh2023cold}, as these theories are characterized by an extensive set of physical constraints that needs to be preserved (e.g., Gauss' law in quantum electrodynamics)~\cite{halimeh2021gauge,halimeh2022stabilizing,rajput2023quantum}). 

Further along the road, the proposed approach of including the energy penalties directly by using only a very small number of ancilla qudits, and thus avoiding the incorporation of slack variables into the energy functions, may be an enabling step for approaching realistic industry-scale and fundamental science problems with large numbers of inequality constraints.

%%%%%%%%%%%%%%%%%%%%%%%%%%%%%%%%%%%%%%%%%%%%%%%%%%%%%%%%%%%%%%%%%%%%
\section{Acknowledgements}
We acknowledge fruitful discussions with Gopal Chandra Santra, Linus Ekstr\o m, and Mikel Garcia de Andoin.
A.B.\ acknowledges funding from the Honda Research Institute Europe. 
S.S.\ and P.H.\ acknowledge funding by the European Union under Horizon Europe Programme, Grant Agreement 101080086 — NeQST.
This project has received funding from the Italian Ministry of University and Research (MUR) through the FARE grant for the project DAVNE (Grant R20PEX7Y3A), was supported by the Provincia Autonoma di Trento, and Q@TN, the joint lab between University of Trento, FBK—Fondazione Bruno Kessler, INFN—National Institute for Nuclear Physics, and CNR—National Research Council.
Project funded under the National Recovery and Resilience Plan (NRRP), Mission 4 Component 2 Investment 1.4 - Call for tender No. 1031 of 17/06/2022 of Italian Ministry for University and Research funded by the European Union – NextGenerationEU (proj.\ nr.\ CN\_00000013).
Project DYNAMITE QUANTERA2\_00056 funded by the Ministry of University and Research through the ERANET COFUND QuantERA II – 2021 call and co-funded by the European Union (H2020, GA No 101017733). 
Views and opinions expressed are however those of the author(s) only and do not necessarily reflect those of the European Commission, the European Union or of the Ministry of University and Research. Neither the European Union nor the granting authority can be held responsible for them.
%%%%%%%%%%%%%%%%%%%%%%%%%%%%%%%%%%%%%%%%%%%%%%%%%%%%%%%%%%%%%%%%%%%%
% \bibliographystyle{plain}
%\bibliographystyle{plainnat}
% \bibliographystyle{quantum}
% \bibliographystyle{apsrev4-2}

%apsrev4-2.bst 2019-01-14 (MD) hand-edited version of apsrev4-1.bst
%Control: key (0)
%Control: author (72) initials jnrlst
%Control: editor formatted (1) identically to author
%Control: production of article title (-1) disabled
%Control: page (0) single
%Control: year (1) truncated
%Control: production of eprint (0) enabled
%

%\bibliography{bib}

\appendix
%%%%%%%%%%%%%%%%%%%%%%%%%%%%%%%%%%%%%%%%%%%%%%%%%%%%%%%%%%%%%%%%%%%%
\newcommand{\I}{\ensuremath{\mathcal{I} } }
\newcommand{\NI}{\ensuremath{N_\mathcal{I} } }

\section{Penalties for Hamming weight constraints}\label{sec: cnots}
This appendix shows how to implement unitaries of the Hamiltonians \eqref{eq: hamiltonian penalties} using a qudit ancilla. 
There exist ways to exponentiate arbitrary boolean functions in quantum computers. In general, these require computing the Fourier transform of the function that needs to be exponentiated. 
In principle, this allows to implement diagonal unitaries of the form of Eq.~\eqref{eq:phasesGr}, albeit with unfavorable, typically exponential, scaling in the number of gates (see, e.g.,  \cite{welch2014efficient,hadfield2021representation}) or with a problem-dependent reduced gate count.

Here, we provide an alternative way to implement such a unitary using only one ancilla qudit and without the need to compute the  Fourier transform. 
We assume a unitary $U_g$ that is parametrized by a function $g(m)$ that only depends on the total Hamming weight (or magnetization)
\begin{equation}
m(\x)=\sum_{i=1}^N x_i
\end{equation}
of the $N$-qubit basis state 
\begin{align}
    \ket{\x}=\ket{x_1,x_2,\dots,x_N}\quad(x_i\in\{0,1\})\:.
\end{align}
This is a common constraint that appears in many scheduling problems such as the EV charging problem. 

For a $N$ qubit state, the Hamming weight can take the $N+1$ integer values $m\in\{0,1,\dots,N\}$, and the function $g(m)$ can have any functional dependence on $m$. 
For our purposes of implementing penalty terms, we assume that it is non-negative for a given subset $ \mathcal{I}$ of  all possible values for $m$, 
\begin{align}
g(m)>0\text{    for    }m\in \mathcal{I}\:.    
\end{align}
For the other states, the function is taken to be zero to indicate feasible states, i.e., $g(m)=0$ for  $m\not\in \mathcal{I}$.
The number of infeasible values for $m$ is denoted by $\NI= |\mathcal{I}|$.

The construction to implement such a function is valid if the action of the unitary $U_g$ can be implemented in a controlled way, where the control state is the ancilla qudit,
\begin{align}
    \ket{y}_a\text{    with  }  y\in\{0,1,\dots,N\}\:.
\end{align}
The  dimension of the ancilla qudit is determined by the number of qubits, $N+1$.

\begin{figure}
    \centering
    \includegraphics[width = \linewidth]{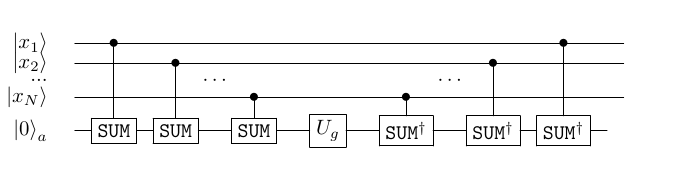}
    \caption{Circuit showing how an arbitrary phase $g(m)$ as function of the magnetization $m(\x)=\sum_ix_i$ can be applied on the qubit register using one ancilla qudit $\ket{y}_a$.}\label{fig: penalty circuit} 
\end{figure}

The circuit implementing the proposed approach is depicted in Fig.~\ref{fig: penalty circuit}.
We start from a product state of the problem qubits, which can be in an arbitrary state, and the ancilla qudit, which is initialized in $\ket{0}_a$, i.e., 
\begin{align}
    \ket{\psi}\ket{0}_a = \sum_{\x} c_{\x} \ket{\x}\ket{0}_a\:.
\end{align}
Then, we  apply a \texttt{SUM} gate that adds the boolean value corresponding to one of the qubits onto the target ancilla.
Repeating this operation for each qubit, we end up in a state where the ancilla qudit has the value $m=m(\boldsymbol{x})=\sum_i x_i$,
\begin{equation}
    \prod_i\texttt{SUM}_{i\to a}\ket{\psi}\ket{0}_a=\sum_{\x} c_{\x} \ket{\x}\ket{m}_a \ .
\end{equation}

After that, we apply a simple phase shift to the ancilla qudit
such that $U_g\ket{m}=e^{ig(m)}\ket{m}$. This operation takes the state to
\begin{align}
   & \sum_{\substack{\x \\ m(\x)\not\in \mathcal{I} } }  c_{\x} \ket{\x}U_g\ket{m}_a + \sum_{\substack{\x\\ m(\x)\in \mathcal{I} } }  c_{\x}\ket{\x} U_g\ket{m}_a  \\
    &=\sum_{\substack{\x \\ m(\x)\not\in \mathcal{I} } }  c_{\x} \ket{\x}\ket{m}_a + \sum_{\substack{\x\\ m(\x)\in \mathcal{I} } }  c_{\x}e^{ig(m)}\ket{\x}\ket{m}_a \:.
\end{align}
After undoing the \texttt{SUM} gates, the final state is 
\begin{equation}
\bigg(\sum_{\substack{\x \\ m(\x)\not\in \mathcal{I} } }  c_{\x} \ket{\x} + \sum_{\substack{\x\\ m(\x)\in \mathcal{I}} }  c_{\x}e^{ig(m)}\ket{\x}\bigg)\ket{0}_a \,,
\end{equation}
which realizes the desired application of the unitary parameterized by a function $g(m)$ on the qubit quantum state $\ket{\psi}$. 
For the case of the constraints we consider, $g$ has the form 
\begin{align}
g(m)= \Theta(m-m_0)\,(m-m_0)^a \:.    
\end{align}
and the corresponding final state is 
\begin{equation}
    \Big(
    \sum_{\substack{\x \\ m(\x)\not\in\ \mathcal{I} } } c_{\x} \ket{\x} + \sum_{\substack{\x\\ m(\x)\in \mathcal{I} }} c_{\x} e^{i(m(\x)-m_o)^a} \ket{\x}
        \Big)\ket{0}_a
\end{equation}
which adds a phase to those states  with $m\in\mathcal{I}=\{m_0+1,m_0+2,\dots,N\}$. 

The entire procedure for arbitrary $g(m)$ requires only $2N$ qudit controlled \texttt{SUM} gates and a single qudit phase shift unitary. 
A requirement for it to work is the availability of qudits with at least $N$ levels, where $N$ is the number of qubits involved in the constraint. This procedure becomes thus particularly favourable when many constraints acting on restricted sets of qubits are present. 

The above procedure can also be generalized to penalties which depend on functions of $m^*$, where     
\begin{equation}
    m^*(\x) = \sum_{i\in N}x_i+\sum_{i\in \Bar{N}}\Bar{x}_i\,.
\end{equation}
$N$ and $\Bar{N}$ are two distinct subsets of the system qubits. 
One needs only to inlclude a $\pi$-rotation ($\sigma_x$) for each qubit belonging to set $\bar{N}$ before (after) the \texttt{SUM}-gate ($\texttt{SUM}^\dagger$-gate).
%%%%%%%%%%%%%%%%%%%%%%%%%%%%%%%%%%%%%%%%%%%%%%%%%%%%%%%%%%%%%%%%%%%%

\end{document}